\documentclass[10pt,twoside,twocolumn]{book}

\usepackage[T1]{fontenc}
\usepackage[latin1]{inputenc}
\usepackage{geometry}
\usepackage{amssymb,amsmath,amsfonts}
\usepackage{nicefrac}
\usepackage{latexsym}
\usepackage{fancyhdr}
\usepackage{graphicx}
\usepackage{amsthm,multibox}
\usepackage{color}
\usepackage{hhline,xspace,rotate}
\usepackage{epsfig}
\usepackage{pst-all}
\usepackage{pst-poly}
\usepackage{pst-coil}
\usepackage{multido}
\usepackage{url}
\usepackage{colortbl}
\usepackage{verbatim}
\usepackage{caption}
\usepackage{subcaption}
\usepackage{braket}

\usepackage{import}
\usepackage{setspace}
\usepackage[breaklinks=true,debug=true]{hyperref}
\makeatletter

\newcommand \thearabicpart {\@arabic\c@part} 

\newlength{\picheight}
\newlength{\titlewidth}
\newsavebox{\picbox}
\savebox{\picbox}{%
  \parbox[b][3.6cm][c]{2.4cm}{}%
}

\def\institute#1{\gdef\@institute{#1}}
\def\@institute{\@latex@warning@no@line{No \noexpand\institute given}}
\def\maketitle{%
\refstepcounter{chapter}%
\twocolumn[
\null
\setlength{\titlewidth}{\textwidth}
\addtolength{\titlewidth}{-2\wd\picbox}
\addtolength{\titlewidth}{-3em}
\hskip\wd\picbox%
\hfill{\large\bf
  \parbox[b]
{\titlewidth}{%
    \centering\@title\\[2ex plus 2 fill]
    \Large\@author\\[0ex plus 1 fill]
    \large\it\@institute%
  }%
}%
\hfill\usebox\picbox
\par%
\vskip 2em%
]%
\thispagestyle{plain}%
\addcontentsline{toc}{chapter}{{\let\\\protect\xspace\@author: \itshape\@title}}%
\setcounter{section}{0}%
}
\newenvironment{abstract}{\section*{Abstract}}{}

\def\@part[#1]#2{%
    \ifnum \c@secnumdepth >-2\relax
      \refstepcounter{part}%
      \addcontentsline{toc}{part}{\thepart\hspace{1em}#1}%
    \else
      \addcontentsline{toc}{part}{#1}%
    \fi
    \def\rjcpartname{#1}
    \markboth{}{}%
    {\centering
     \interlinepenalty \@M
     \normalfont
     \ifnum \c@secnumdepth >-2\relax
       \huge\bfseries\sffamily \partname\nobreakspace\thepart
       \par
       \vskip 20\p@
     \fi
     \Huge \bfseries #2\par}%
    \@endpart}

\def\@lbibitem[#1]#2{%
  \@skiphyperreftrue
  \H@item[%
    \ifx\Hy@raisedlink\@empty
      \hyper@anchorstart{cite.\thechapter.#2}\@BIBLABEL{#1}\hyper@anchorend
    \else
      \Hy@raisedlink{\hyper@anchorstart{cite.\thechapter.#2}\hyper@anchorend}%
      \@BIBLABEL{#1}%
    \fi
    \hfill
  ]%
  \@skiphyperreffalse
  \if@filesw
    \begingroup
      \let\protect\noexpand
      \immediate\write\@auxout{%
        \string\bibcite{\thechapter.#2}{#1}%
      }%
    \endgroup
  \fi
  \ignorespaces
}%
\def\@bibitem#1{%
  \@skiphyperreftrue\H@item\@skiphyperreffalse
  \Hy@raisedlink{\hyper@anchorstart{cite.\thechapter.#1}\relax\hyper@anchorend}%
  \if@filesw
    \begingroup
      \let\protect\noexpand
      \immediate\write\@auxout{%
        \string\bibcite{\thechapter.#1}{\the\value{\@listctr}}%
      }%
    \endgroup
  \fi
  \ignorespaces
}%
\def\@citex[#1]#2{%
  \let\@citea\@empty
  \@cite{\@for\@citeb:=#2\do
    {\@citea\def\@citea{,\penalty\@m\ }%
     \edef\@citeb{\thechapter.\expandafter\@firstofone\@citeb\@empty}%
     \if@filesw\immediate\write\@auxout{\string\citation{\@citeb}}\fi
     \@ifundefined{b@\@citeb}{\mbox{\reset@font\bfseries ?}%
       \G@refundefinedtrue
       \@latex@warning
         {Citation `\@citeb' on page \thepage \space undefined}}%
       {\hbox{\csname b@\@citeb\endcsname}}}}{#1}}

\renewenvironment{thebibliography}[1]
     {\section*{\refname}%
      \@mkboth{\MakeUppercase\refname}{\MakeUppercase\refname}%
      \list{\@biblabel{\@arabic\c@enumiv}}%
           {\settowidth\labelwidth{\@biblabel{#1}}%
            \leftmargin\labelwidth
            \advance\leftmargin\labelsep
            \@openbib@code
            \usecounter{enumiv}%
            \let\p@enumiv\@empty
            \renewcommand\theenumiv{\@arabic\c@enumiv}}%
      \sloppy
      \clubpenalty4000
      \@clubpenalty \clubpenalty
      \widowpenalty4000%
      \sfcode`\.\@m}
     {\def\@noitemerr
       {\@latex@warning{Empty `thebibliography' environment}}%
      \endlist}

\makeatother

\usepackage[english]{babel}

\begin{document}

\institute{CEA-Saclay, IRFU/SPP \\ email: antoine.chapelain@cea.fr}

\title{Charge asymmetry of top quark-antiquark pairs}

\author{Antoine Chapelain }

\maketitle

\begin{abstract}

In this note we present the charge asymmetry measurements of top quark-antiquark pairs at hadron colliders.

\end{abstract}

\section{Introduction}

Among the known twelve fermions, which are the fundamental bricks of matter, the top quark is the
latest to have been discovered at the Tevatron Fermilab Collider near Chicago by the CDF and D0 experiments in 1995~\cite{tev_top_discovery}.
The top quark was found to be the heaviest particle ever observed. Its mass is about the mass of the gold atom, which
is extremely heavy for a point-like particle.
Due to its large mass, studying the top quark could be a window towards so-called new physics, $i.e.$, physics that lies beyond the Standard Model of particle physics.
The top quark can be scrutinized at hadron colliders since the Tevatron and LHC colliders produced numerous top quark-antiquark~\footnote{Each fermion has a corresponding antifermion,
which have the same properties but opposite electric charges.} pairs.
The charge asymmetry is one of the properties of top quark-antiquark pairs.
Indeed the strong interaction predicts that when produced through quark-antiquark collisions the top quark and antiquark are not produced isotropically. 
The top quark is preferentially produced in the direction of the incoming quark while the top antiquark is preferentially produced in the direction of the incoming
antiquark in the incoming quark-antiquark rest frame. As the top quark and antiquark have opposite electric charge, it will result in a charge asymmetry (excess of positive/negative charge
in the incoming quark/antiquark direction).

\begin{figure}[htb]
\begin{center}\includegraphics[%
  width=4.2cm,
  keepaspectratio]{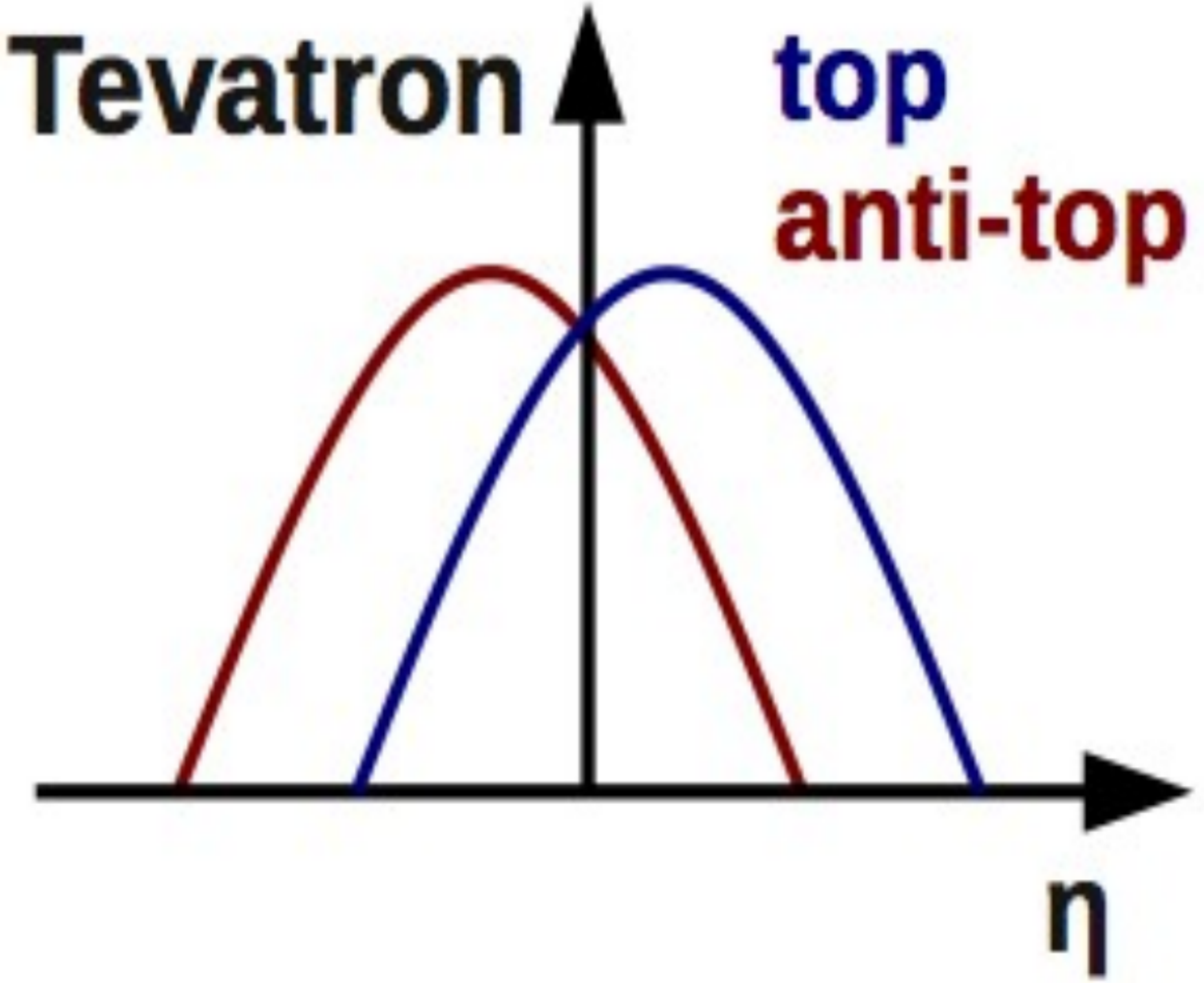}
\includegraphics[%
  width=4.2cm,
  keepaspectratio]{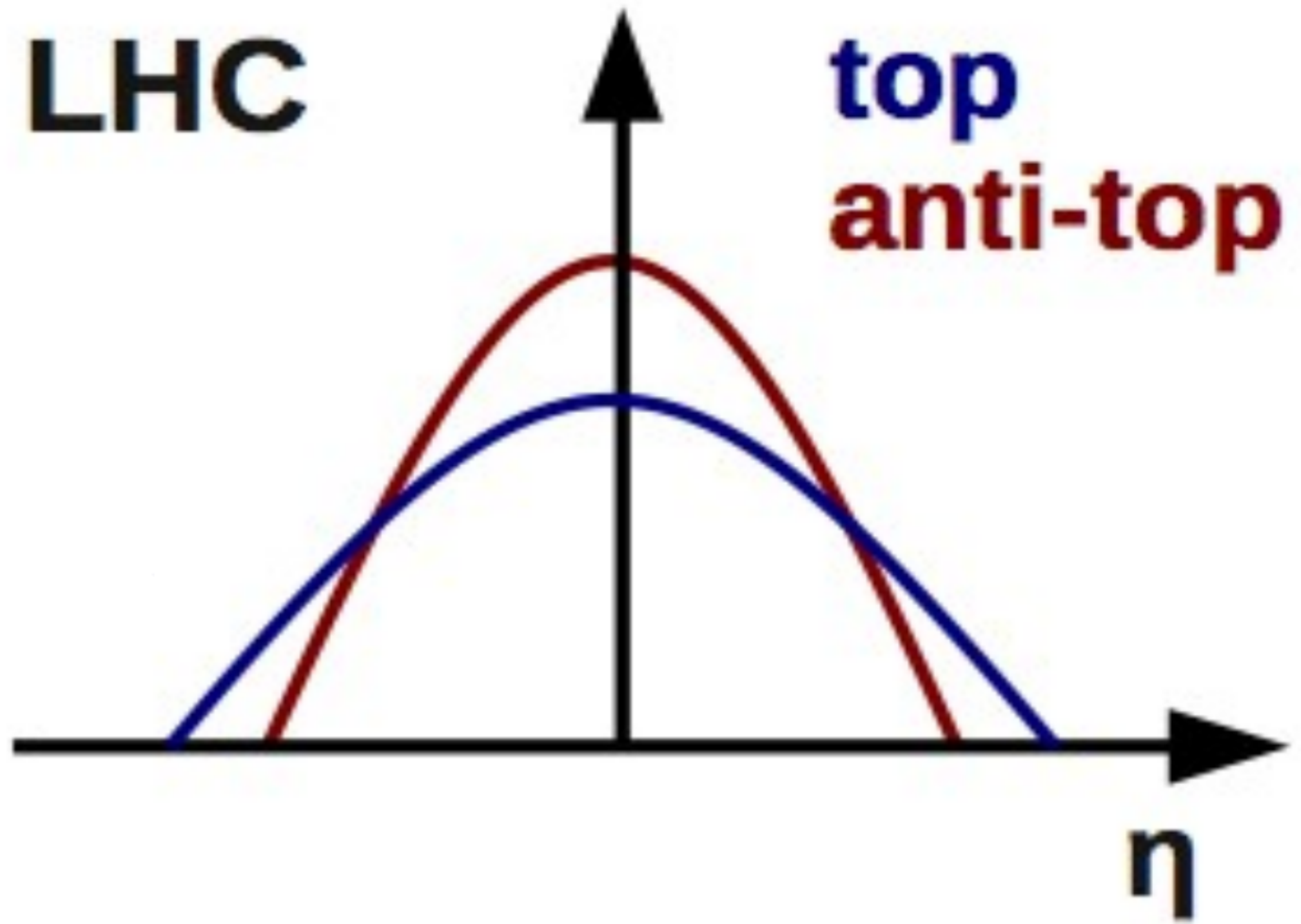}\end{center}
\caption{Rapidity distributions of the top quark and antiquark at the Tevatron (top) and LHC (bottom).}
\label{fig:asym}
\end{figure}

To quantify this effect we use the rapidity $y$ \mbox{(or pseudorapidity $\eta$)}. It is defined approximatively as \mbox{$y \simeq -ln(tan(\theta/2))$} where $\theta$ is the angle
between the flight direction of the top quark (antiquark) and the beam direction.
Figure~\ref{fig:asym} shows how the charge asymmetry appears at the Tevatron and LHC. At the Tevatron, which is a proton-antiproton collider,
there is a forward-backward (or right-left) asymmetry since the incoming quark-antiquark collision frame is almost equal to the quark-antiquark rest frame.
Since the LHC is a proton-proton collider the quarks carry on average a higher momentum than the antiquarks which come from the sea of the proton. The top quark will be thus
emitted more forward or backward and the top antiquark will be emitted more central. Measuring the charge asymmetry at the Tevatron and LHC is therefore
complementary.

The charge asymmetry can also be measured directly using the leptons coming from the decay of the top quark and antiquark since the flight direction of these
leptons is correlated with the flight direction of the top quark/antiquark. This measurement is simpler because the flight direction of the leptons is
directly measured in the detector while the top quark flight direction need to be reconstructed from the decay products of the top quark.

\begin{figure}[htb]
\begin{center}\includegraphics[%
  width=7cm,
  keepaspectratio]{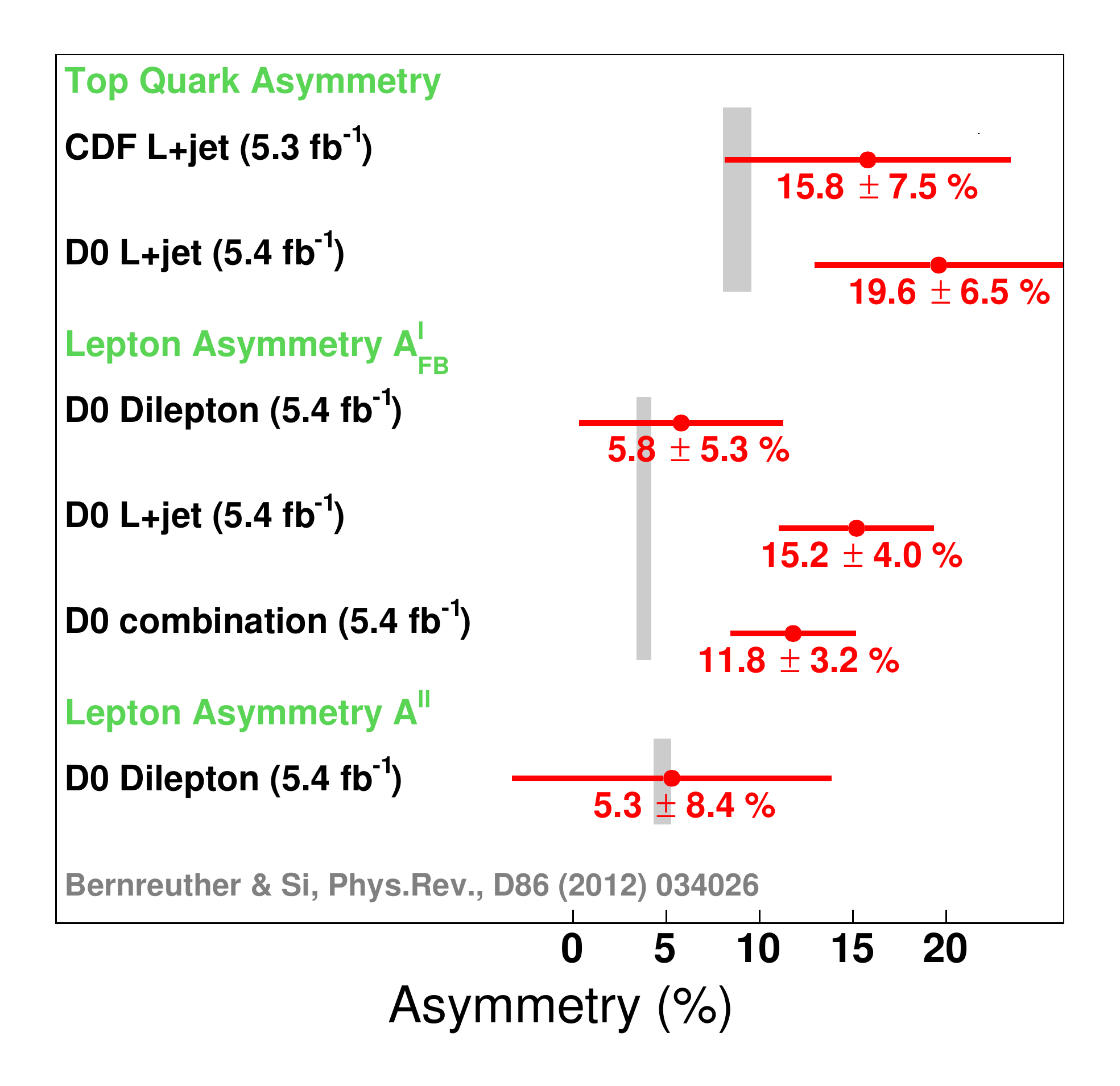}\end{center}
\caption{Summary of the asymmetry measurements at the Tevatron in 2011.}
\label{fig:2011}
\end{figure}

In 2011 the CDF and D0 collaborations reported measurements higher than the predictions as summarized in Figure~\ref{fig:2011}.
The Tevatron stopped data taking in September 2011. Updating the asymmetry measurement with the full CDF and D0 recorded dataset is underway.

In Sec. 2 we will focus on the measurement performed at D0 in the dilepton channel~\cite{d0_dilepton} and in Sec. 3 on the measurement performed at ATLAS in the dilepton channel at 8 TeV.
Section 4 summarizes the inclusive charge asymmetry measurements at hadron colliders.

\section{Dilepton measurement at D0}

\begin{figure}[htb]
\begin{center}\includegraphics[%
  width=5cm,
  keepaspectratio]{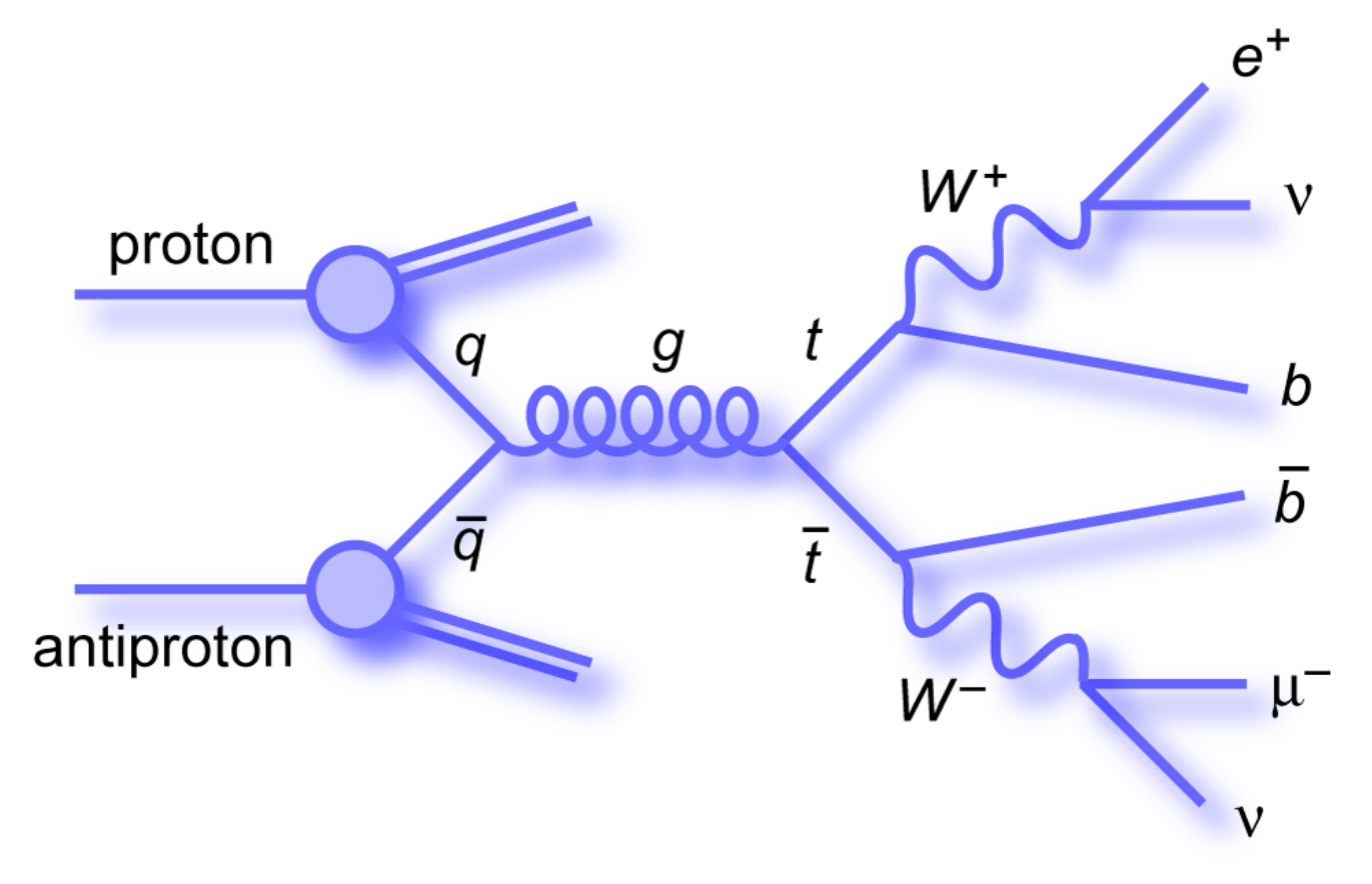}\end{center}
\caption{Topology of a dilepton $t\bar{t}$ event produced through quark-antiquark annihilation at the Tevatron.}
\label{fig:ttbar_topo}
\end{figure}

The $t\bar{t}$ dilepton final state (see Fig.~\ref{fig:ttbar_topo}), or dilepton channel, is characterized by two leptons with opposite electric charge, at least two jets coming from
the two $b$-quarks, and missing energy due to the two neutrinos escaping the detector (the neutrinos and the charged leptons are both coming from the decay of the $W$ bosons). 

\begin{figure}[htb]
\begin{center}\includegraphics[%
  width=5cm,
  keepaspectratio]{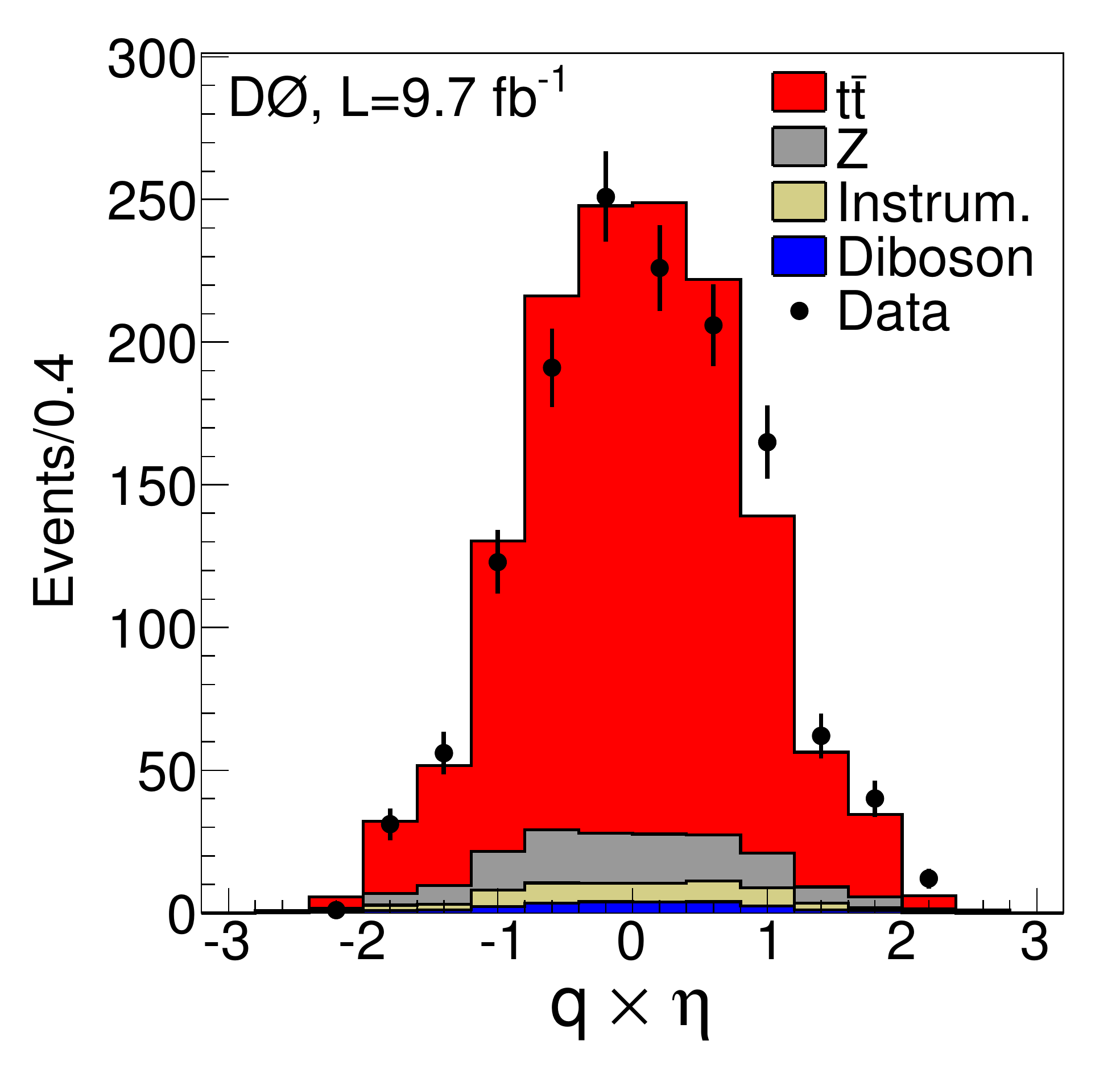}
\includegraphics[%
  width=5cm,
  keepaspectratio]{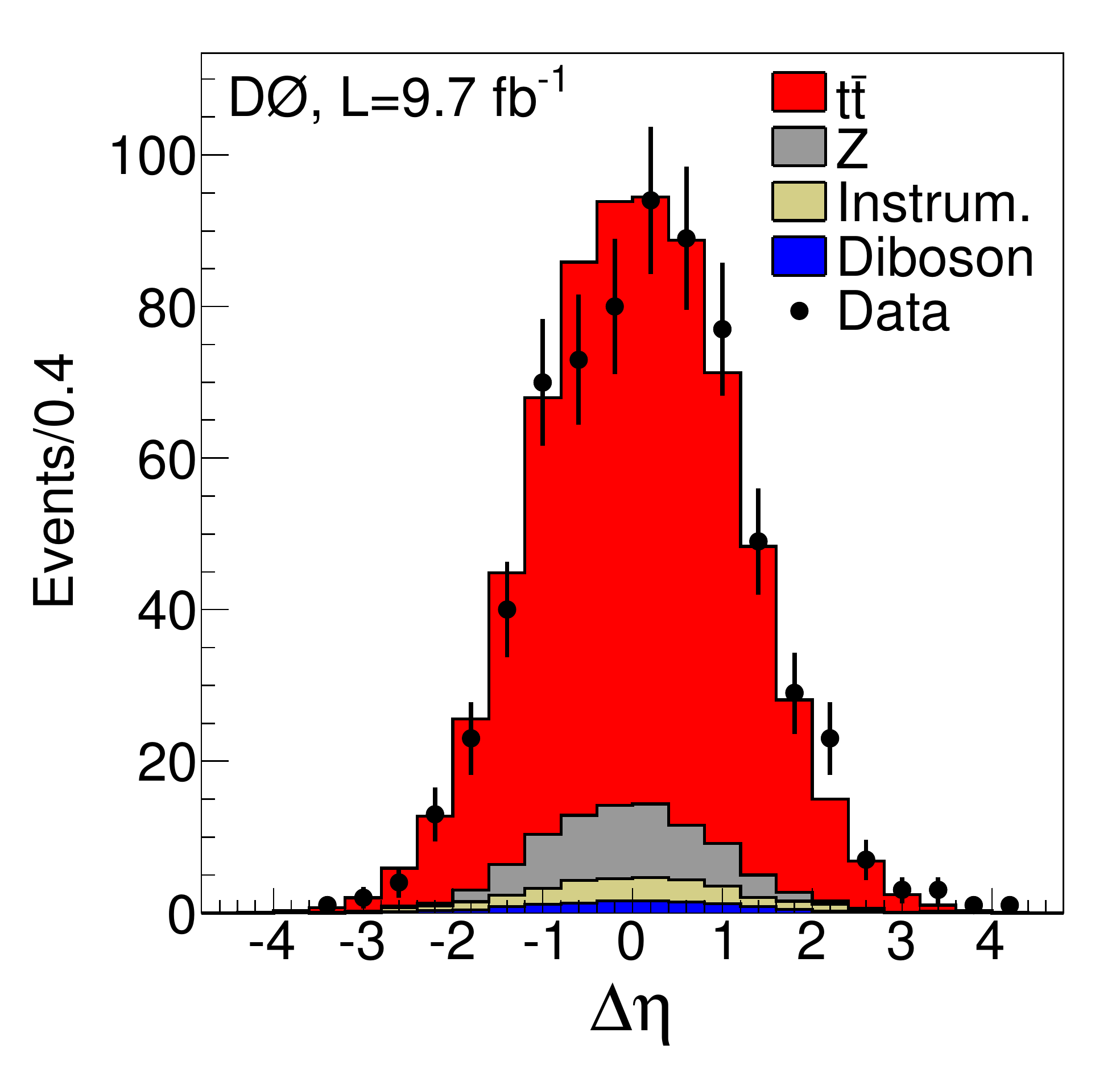}\end{center}
\caption{Rapidity distributions used to compute the asymmetry in the dilepton channel at D0~\cite{d0_dilepton}.}
\label{fig:chapelain_dist_fiducial}
\end{figure}

The dilepton channel suffers from small statistics because of a small branching ratio but on the other hand have a small amount of background.
The measurement of the forward-backward asymmetries through leptons is performed using the two distributions in Fig.~\ref{fig:chapelain_dist_fiducial}.
The single-lepton $A^\ell_{FB}$ asymmetry is defined with the $q\times\eta$ distribution looking at each lepton independently if the lepton goes in the forward ($\eta>0$)
or backward ($\eta<0$) direction. 
The $\Delta\eta$ distribution built as the difference of lepton pseudorapidities is used to measure the lepton-pair $A^{\ell\ell}$ asymmetry.
$A^\ell_{FB}$ and $A^{\ell\ell}$ are computed as the relative difference between the forward and backward region of the relevant distributions using the data from which
we subtracted the expected background.

\begin{figure}[htb]
\begin{center}\includegraphics[%
  width=7cm,
  keepaspectratio]{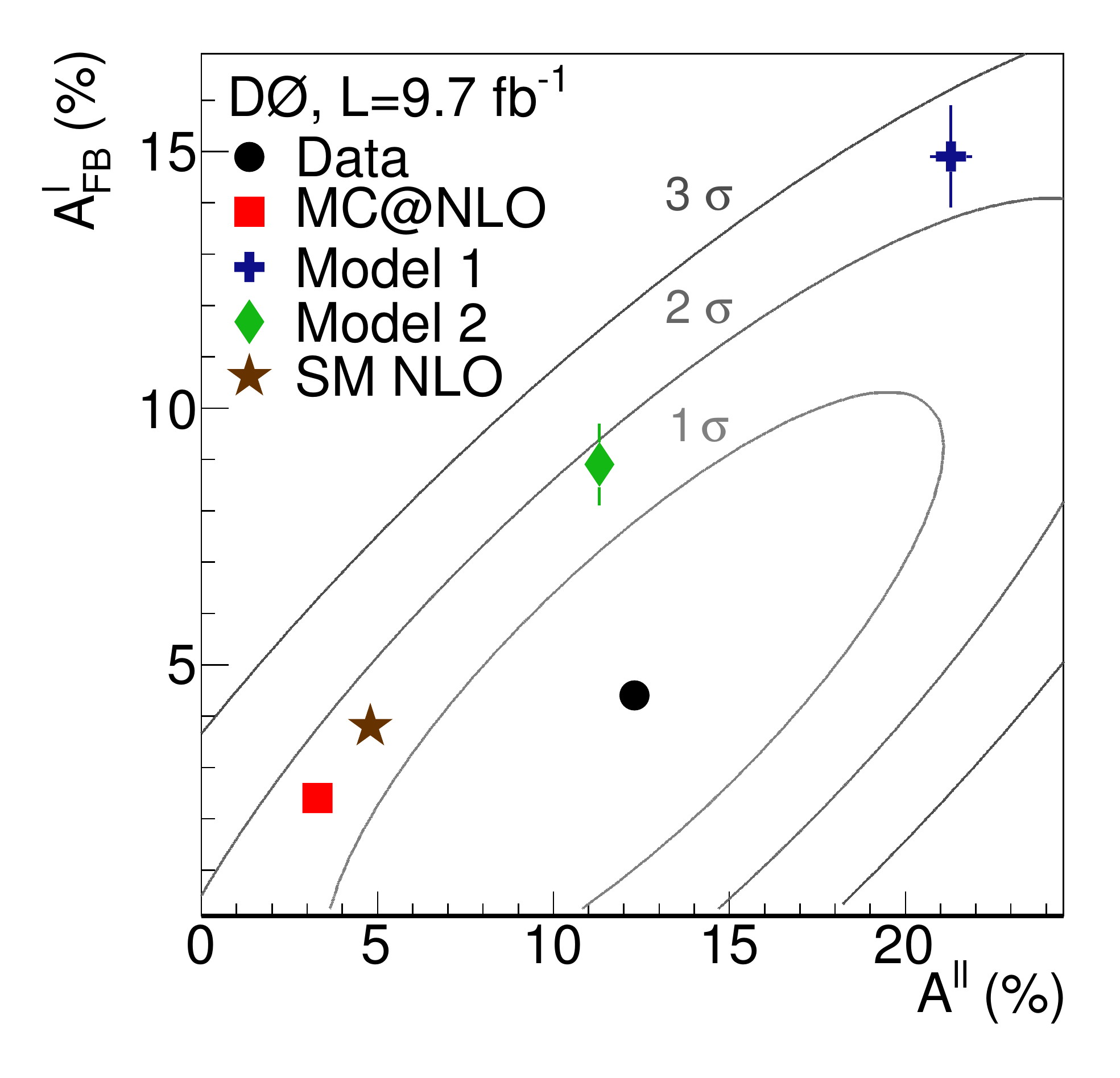}\end{center}
\caption{$A^{\ell}_{FB}$ versus $A^{\ell\ell}$ in the dilepton channel at D0~\cite{d0_dilepton}. The black dot represents the measurement with the uncertainty ellipses
corresponding to 1, 2 and 3 standard deviation. ``SM NLO'' represents the most recent theoretical prediction, ``MC@NLO'' is the event generator used to simulated
the $t\bar{t}$ signal and ``Model 1'' and ``Model 2'' are two new physics model that could explain the 2011 observed tension at the Tevatron between measurements
and predictions.}
\label{fig:chapelain_al_vs_all}
\end{figure}

\begin{figure}[htb]
\begin{center}\includegraphics[%
  width=7cm,
  keepaspectratio]{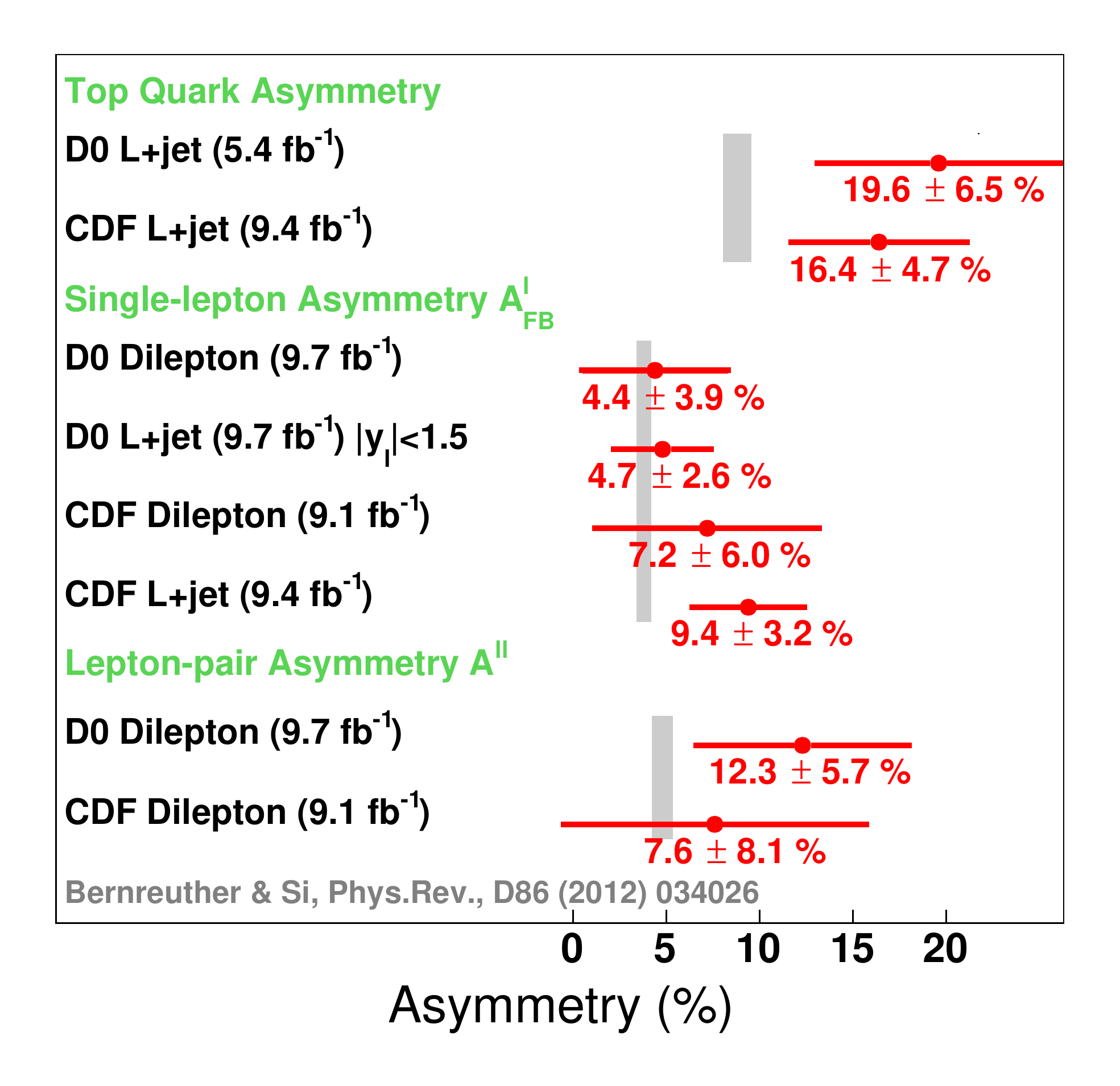}\end{center}
\caption{Summary of the asymmetry measurement at the Tevatron in 2013.}
\label{fig:2013}
\end{figure}

In Fig.~\ref{fig:chapelain_dist_fiducial} the black dots represent the data, the colored histograms represent the predictions: the $t\bar{t}$ signal in red,
and the different backgrounds in grey, yellow and blue. At this level we performed the asymmetry measurements in the detector, $i.e.$, 
distorted by detector effects that need to be corrected for.
We first correct for the selection efficiency, $i.e.$, for the fact that we do not observe all the produced dilepton $t\bar{t}$ events in the detector.
We then correct for the limited spatial coverage of the detector. Once these corrections are made we can compare the measurements to the theoretical predictions. 
Table~\ref{tab:chapelain_d0_results} shows the measured and predicted values. Both are agreement within the uncertainties.
It is interesting to look at the ratio $A^\ell_{FB}/A^{\ell\ell}$ since the two asymmetries are strongly correlated and because the systematic uncertainty is reduced due to
cancellations. Figure~\ref{fig:chapelain_al_vs_all} shows
the measured value in black together with different predictions. The measured value of $0.36 \pm 0.20$ is in agreement with the prediction of $0.79 \pm 0.10$ at the level of two
standard deviations.
Figure~\ref{fig:2013} summarizes all the current measurements at the Tevatron. We can see that the tensions between measurements and predictions observed in 2011 (see Fig.~\ref{fig:2011}) vanished. 
Two measurements from D0 have still to be released in 2014. The focus is now on the CDF-D0 combination of these different results to achieve the best possible precision. 

\begin{table}[hbt]
\begin{center}
\begin{tabular}{lcc}
\hline\hline
  & Measured & Predicted \\
\hline
$A^\ell_{FB}$ & $\phantom{0}4.4 \pm 3.7 \pm 1.1$ & $3.8 \pm 0.3$ \\ 
$A^{\ell\ell}$ & $12.3 \pm 5.3 \pm 1.5$ & $4.8 \pm 0.4$ \\
\hline\hline
\end{tabular}
\end{center}
\caption{Measured and predicted values of the two leptonic asymmetries
in the D0 dilepton channel. The first uncertainty on the measured values
is statistical and the second is systematic.}
\label{tab:chapelain_d0_results}
\end{table} 

\section{Dilepton measurement at ATLAS}

\begin{figure}[htb]
\begin{center}\includegraphics[%
  width=5cm,
  keepaspectratio]{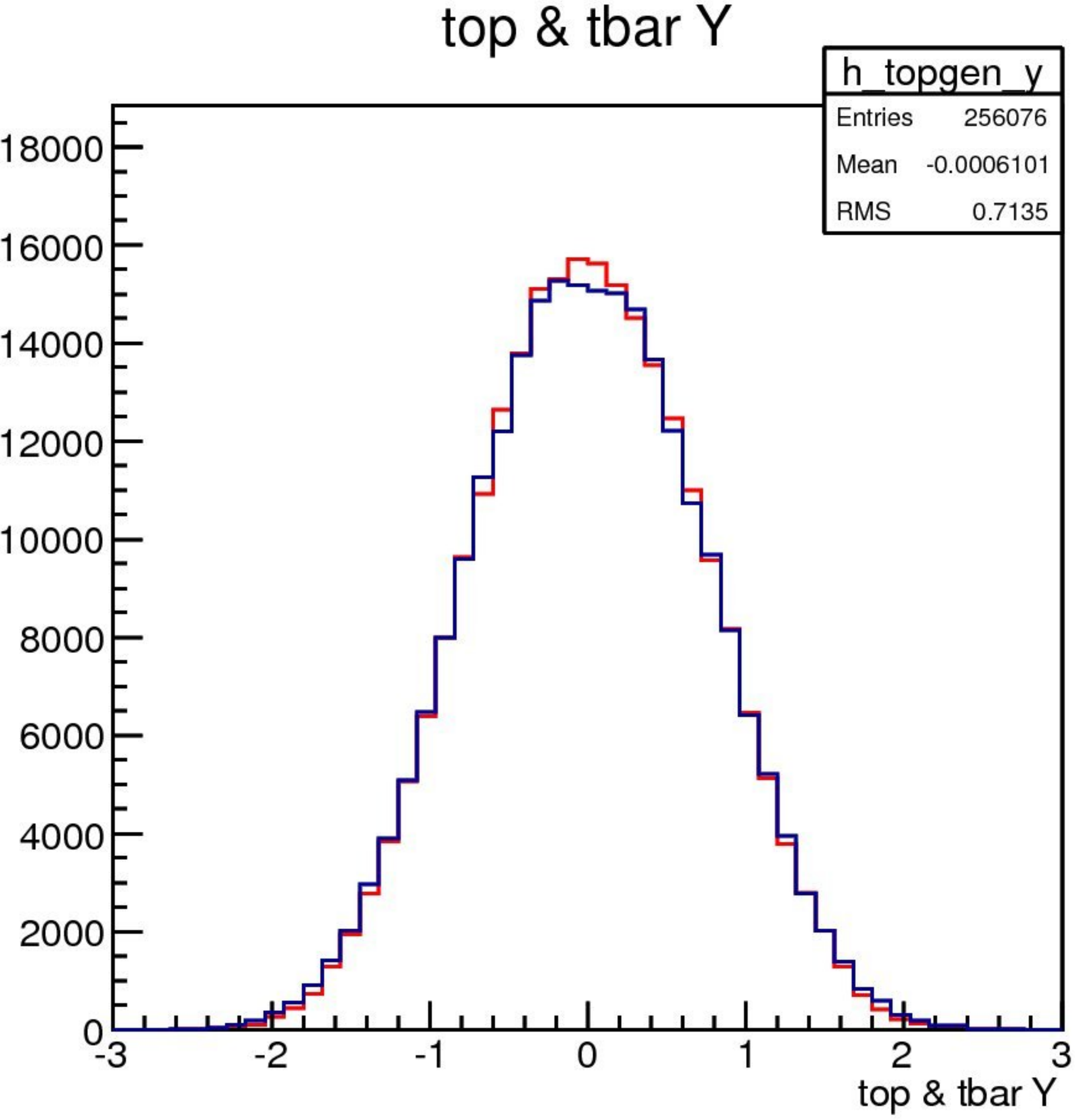}
\end{center}
\caption{Rapidity distributions at truth (red) and reconstructed (blue) level.}
\label{fig:chapelain_dy}
\end{figure}

As explained earlier, measuring the charge asymmetry at the LHC and the Tevatron is complementary. This section is focusing on the charge asymmetry measurement in the dilepton
channel at ATLAS. Both the asymmetry of the lepton coming from the top quark/antiquark and of the $t\bar{t}$ pairs are measured.
The top quark is not directly observed in the detector due to its very short lifetime ($10^{-23}~s$). Thus we need to reconstruct its kinematic from its observed decay products.
To do so we use the energy and momentum conversation at each decay vertices of the decay chain. 
We obtain then a system of 16 equations and 22 unknowns which cannot be solved. Making several assumptions and fixing the masses of the $W$ bosons and the top quarks to their measured values
we finally end up with 18 equations and 18 unknowns.
For a given event we obtain several solutions. We define a weight for each solution according to its probability to be a $t\bar{t}$ event. This probability is computed using
the matrix element of the $gg \rightarrow t\bar{t}$ process. The solution with the highest weight is selected. This method is called the ``Matrix Element method''~\cite{ME}.
We use the simulation to test the performances of this reconstruction method. The variable we are interested in to compute the asymmetry is the rapidity $y$ of the top quark and antiquark. 
Figure~\ref{fig:chapelain_dy} shows the $y$ distribution at the so-called ``truth'' level and after reconstruction.
The truth level is what is generated with the simulation and the reconstructed level is what we reconstruct after the simulation of physics and detector effects. 
We see that the reconstructed distribution reproduces the behavior of the truth distribution well. 

The asymmetry is computed using the observable defined as $\Delta|y|=|y_{top}|-|y_{antitop}|$.
In the example of Fig.~\ref{fig:chapelain_dy} we are able to reconstruct the correct sign of $\Delta|y|$ in 70~\% of the cases. This performance is rather satisfying
and very similar to performances of other reconstruction method. This measurement at 8 TeV with the ATLAS detector is still ongoing and will be released soon.

\section{Current status and conclusion}

\begin{figure}[htb]
\begin{center}\includegraphics[%
  width=7cm,
  keepaspectratio]{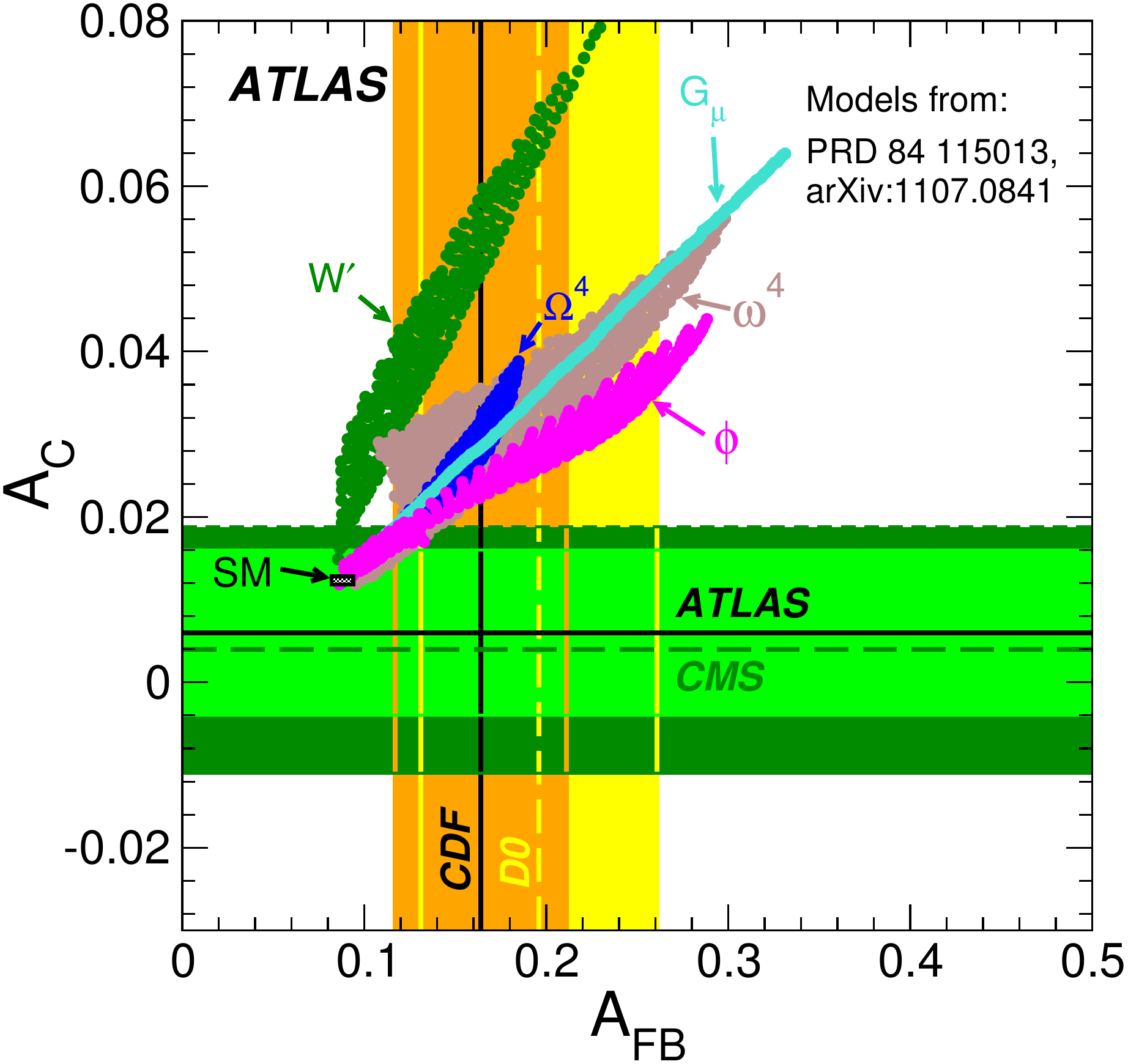}
\end{center}
\caption{Summary of the measurements at the Tevatron and LHC and predictions from different physics models~\cite{ATLAS_NP}.}
\label{fig:sum_np}
\end{figure}

The Tevatron and LHC are both the most powerful proton-antiproton and proton-proton colliders, respectively.
They allow to conduct complementary studies on the charge asymmetry of the top quark-antiquark pairs.
In 2011 the Tevatron measurements showed tension between measurements and predictions. The latest results with the full
statistics recorded by the CDF and D0 experiments tend to indicate a better agreement between predictions and
measurements. At the LHC, so far all the measurements are in good agreement with the predictions. 
Some physics model beyond the Standard Model could explain the deviations observed at the Tevatron in 2011 while still in agreement
with the observation at the LHC (see Fig.~\ref{fig:sum_np}).
We can see on Fig.~\ref{fig:sum_np} that a small region of phase space is still allowed for these new physics model. The new results from D0, as well as new
results from ATLAS and CMS are expected to be able to make a conclusive statement. The year 2014 is thus very promising
to understand deeper the charge asymmetry of the top quark-antiquark pairs.

\pagestyle{empty}


\begin{thebibliography}{2}
\bibitem{tev_top_discovery} The CDF Collaboration, Phys.Rev.Lett. 74.2626 (1995); The D0 Collaboration, Phys.Rev.Lett. 74.2632 (1995).
\bibitem{d0_dilepton} The D0 Collaboration, Phys.Rev. D88 (2013) 112002.
\bibitem{ME} F. Fiedler, A. Grohsjean, P. Haefner, P. Schieferdecker,  	Nucl.Instrum.Meth. A624 (2010), 203-218.
\bibitem{ATLAS_NP} The ATLAS Collaboration, arXiv:1311.6724 (2013).

\end{thebibliography}
\end{document}